\documentclass[Journal,letterpaper,SectionNumbers,InsideFigs,NoLineNumbers]{ascelike-new}

\usepackage[utf8]{inputenc}
\usepackage{xcolor}
\usepackage[T1]{fontenc}
\usepackage{lmodern}
\usepackage{graphicx}
\usepackage[figurename=Fig.,labelfont=bf,labelsep=period]{caption}
\usepackage{subcaption}
\usepackage{amsmath}
\usepackage{adjustbox}
\usepackage{tikz}
\usepackage{enumitem}
\usepackage{multirow}
\usepackage{booktabs}
\usepackage{comment}
\usepackage{multirow}
\usepackage{color}
\usepackage{colortbl}
\usepackage[normalem]{ulem}

\usepackage{newtxtext,newtxmath}
\usepackage[colorlinks=true,citecolor=black,linkcolor=black]{hyperref}

\hypersetup{
    colorlinks = true,
    urlcolor   = black,
    citecolor  = purple,
    linkcolor=black,
}

\NameTag{\small{10.1061/JHEND8/HYENG-14074}}

\DeclareMathOperator\erf{erf}
\usepackage{tikz,pgfplots}
\usepgfplotslibrary{groupplots}
\usepackage{pgfplotstable}
\usepgfplotslibrary{external}
\usetikzlibrary{positioning,arrows.meta,shapes}
\pgfplotsset{compat=1.15}

\usepackage{lipsum}
\usepackage{makecell}
\usepackage{longtable}

\begin{document}


\title{Improving the measurement of air-water flow properties using remote distance sensing technology \\
\normalsize{\normalfont{accepted for publication, ASCE Journal of Hydraulic Engineering}}
}

\author[1]{M. Kramer}
\author[2]{D. B. Bung}

\affil[1]{Senior Lecturer, UNSW Canberra, School of Engineering and Technology (SET), Canberra,
ACT 2610, Australia,  ORCID 0000-0001-5673-2751, Email: m.kramer@unsw.edu.au}

\affil[2]{Professor, Hydraulic Engineering Section (HES), FH Aachen, University of Applied Sciences, Aachen, Germany, ORCID 0000-0001-8057-1193, Email: bung@fh-aachen.de}

\maketitle

\begin{abstract}
In recent years, there has been an increasing research interest in the application of remote sensing technology to highly aerated flows, which is because this technology holds the ultimate promise to enable safe and accurate measurements of real-word air-water flows in natural and human made environments. Despite the increasing number of publications, some fundamental questions, such as ``what do we measure'' or ``what can we measure'', have not been answered conclusively. In this study, we hypothesize that laser distance sensors are able to measure the concentration of entrapped air, which we demonstrate using two seminal air-water flow types, namely a submerged hydraulic jump and flows down a stepped spillway. By converting our free-surface signals into time series of instantaneous air concentrations, we also show that a dual laser triangulation setup enables the extraction of basic air-water flow parameters of the upper flow region, comprising interface count rates, interfacial velocities, and turbulence levels, while we acknowledge that some sensor characteristics, such as beam diameters, can lead to measurement biases. Overall, this study represents a major advancement in the remote measurement of air-water flow properties. Future collective research effort is required to overcome remaining challenges.    

\end{abstract}

\section{Introduction}
High-velocity free-surface flows commonly involve self-aeration, which is due to turbulent forces exceeding gravity and surface tension. For design purposes of hydraulic infrastructure, it is important to consider the effects of self-aeration, such as flow bulking, drag reduction, and enhanced air-water mass transfer. Oftentimes, laboratory investigations are performed at scaled physical models to predict relevant air-water flow parameters, including flow depth, air concentration, interfacial velocity, and turbulence intensity. However, the applicability of classical single-phase flow measurement instrumentation, such as Acoustic Doppler Velocimeters, Laser Doppler Anemometers, or Pitot tubes, is limited in aerated flows, which is because entrained air bubbles hinder the propagation of light and sound waves \cite{Chanson2013}. Most common multiphase flow measurement instruments are intrusive phase-detection probes, which are able to identify phase changes at the sensor's tips by a change of resistivity or light refraction \cite{Neal1963,Cartellier1991}. To date, these sensors are still considered the most reliable instrumentation for air-water flows, despite being relatively expensive (phase-detection optical-fibre probes) or being custom-made, with current design dating back to the 1970s (phase-detection conductivity probes). As such, access to a broader research community is limited and a relatively cheap and robust sensor for measurements of most relevant air-water flow parameters is desirable. 

When performing laboratory-scale experiments of aerated flows, scale effects arise when extra\-po\-lating obtained results to the real-world scale \cite{Pfister2014}, implying that real-world prototype measurements of air-water flows are essential for hydraulic design purposes \cite{Chanson2013,Erpicum2020}.  However, the installation of intrusive flow measurement instrumentation in real-world environments is challenging, with further restrictions arising from limited access, occurrence of three-phase air-water-sediment flows (and potential debris), as well as very high flow velocities. This is highlighted by the small number of prototype studies documented in literature, for example \cite{Cain1978,Bai2021,Hohermuth2021a,Wang2022}. Remote sensing technologies may help to overcome aforementioned difficulties of intrusive devices.

\renewcommand*{\arraystretch}{0.7}
\begin{small}
\begin{table}
\caption{Key publications on remote sensing of aerated flows using distance sensors; ADM = acoustic displacement meter (or ultrasonic sensor); HJ = hydraulic jump; ToF =  Time-of-Flight; LTS = laser triangulation (sensor); FS = free-surface}
\begin{tabular}{c c c c c}
\toprule
Reference & Instrumentation & Principle &  Parameter & Application\\
\midrule
\citeN{BLENKINSOPP20101059} & LIDAR & ToF & $h$ & laboratory breaking wave\\
\citeN{BLENKINSOPP20121}  & LIDAR & ToF & $h$ & laboratory breaking wave\\
\citeN{Bung2013} & ADM & ToF & $h$ & laboratory spillway\\
\citeN{Martins2016}  & LIDAR & ToF & $h$ & field breaking wave\\
\citeN{Valero2016} & ADM & ToF & $h$, $\overline{c}$ & laboratory spillway\\
\citeN{Martins2017}  & LIDAR & ToF & $h$ & field undular bore\\
 \citeN{Rak2017} & LIDAR  & ToF & $h$ &  laboratory confluence\\
 \citeN{Montano2018} & LIDAR & ToF  & $h$ & laboratory HJ \\
\citeN{Kramer2019LIDAR} & LIDAR & ToF & $h$, $\overline{c}$, $\overline{u}_\text{FS}$  & laboratory spillway\\
 \citeN{Rak2020} & LIDAR + camera  & LTS &  $h$ & laboratory confluence \\
  \citeN{Bung2021} & RGB-D camera  & LTS & $h$ & laboratory HJ \\
 \citeN{Cui2022} & ADM  & ToF & $h$, $c$, $\overline{c}$, $F$ & laboratory spillway \\
\multirow{2}{*}{Present study} & Laser point sensor & \multirow{2}{*}{LTS}& $h$, $c$, $\overline{c}$, $F$   & laboratory spillway \\
&  (SICK OD2000) & & $u$, $\overline{u}$, $\overline{u'u'}$ & laboratory HJ \\
\bottomrule
\end{tabular}
\end{table}
\label{table:intro}
\end{small}

In Table \ref{table:intro}, we present a list of key publications on the use of remote sensing/non-intrusive distance sensors in highly-aerated flows, \textcolor{black}{with a specific focus on laser and acoustic displacement sensors}, further including developments in the measurement of various air-water flow parameters, i.e., the flow depth ($h$), air concentration ($c$), interface count rate ($F$), interfacial velocity ($u$), and Reynolds stresses ($\overline{u'u'}$); here, the $\overline{\phantom{c}}$ operator indicates a time-averaged quantity. We note that a recent review article of \citeN{Rak2023} summarises methods for free-surface measurements in aerated flows, including a comprehensive list of light detection and ranging (LIDAR) and laser triangulation applications to supercritical flows, as well as a detailed description of various light reflection mechanisms at air-water interfaces. Importantly, it is discussed that the performance of LIDAR and laser triangulation was comparable, while different data retention rates were observed \cite{Rak2023}. Despite recent advances as per Table \ref{table:intro}, we believe that several aspects have not been sufficiently clarified in past literature, which we shortly summarise as follows:

\begin{enumerate}[label=\roman*)]
\setlength{\itemsep}{0pt}
    \item In aerated flow literature, it has oftentimes been expressed that a ``classical free-surface may not exist'', and that it is more appropriate to refer to so-called mixture flow depths, such as $y_{90} = y(\overline{c}=0.9)$. We note that \citeN{Kramer2024} mathematically defined a free-surface for aerated flows, which however requires the discrimination of entrapped air, i.e., air transported along wave peaks and throughs, and entrained air, in form of air bubbles, both mechanisms first described by \citeN{wilhelms2005bubbles}.
       
    \item The question on ``what do we measure'' with remote sensing devices in highly aerated flows has rarely been addressed systematically. Either, a comparison with intrusive measurements is missing \cite{Rak2017,Montano2018,Rak2020}, or there has been no differentiation between the measurement of entrapped and entrained air \cite{Bung2013,Kramer2019LIDAR,LI2021110392}. We believe that it is necessary to compare air-water surface levels ($h$), measured by a remote sensing device, to intrusively measured mixture flow depths, and to determine whether a measurement instrument detects air-water surfaces or entrained air bubbles.
  
    \item Free-surface deformations and fluctuations are intrinsically connected to the air-concentration distribution, which has been shown by \citeN{Valero2016} and \citeN{Cui2022}.
\end{enumerate}

In this study, we use laser triangulation sensors (LTS) to measure air-water flow properties in two seminal aerated flow types, including a submerged hydraulic jump and a stepped spillway. \textcolor{black}{Note that we selected a submerged hydraulic jump over a classical hydraulic jump for the sake of comparability between different flow measurement instruments, simply because a submerged HJ resembles a more stable flow configuration}. In $\S$ \ref{sec:methods}, we introduce experimental facilities, flow conditions, and instrumentation, followed by a detailed description of the LTS measurements and associated signal processing ($\S$ \ref{sec:laser}). Our results show that ADM and LTS are able to measure \textit{entrapped} air and associated interface count rates ($\S$ \ref{sec:airconcentration}), while our dual-sensor arrangement allowed, for the first time, to remotely measure interfacial velocities and turbulence levels ($\S$ \ref{sec:velocities}). Through the application of a recently introduced two-state superposition approach to our spillway data, we show that different flow measurement instrumentation (ADM, LTS, dipping probe) provide varying information on entrapped air ($\S$ \ref{sec:discussion}). Subsequently, we articulate the need for future systematic investigations on the remote measurement of air-water flow properties by the research community. 

\newpage
\section{Experimental facilities, flow conditions, and instrumentation}
\label{sec:methods}

\begin{figure*}[h!]
\centering
\includegraphics{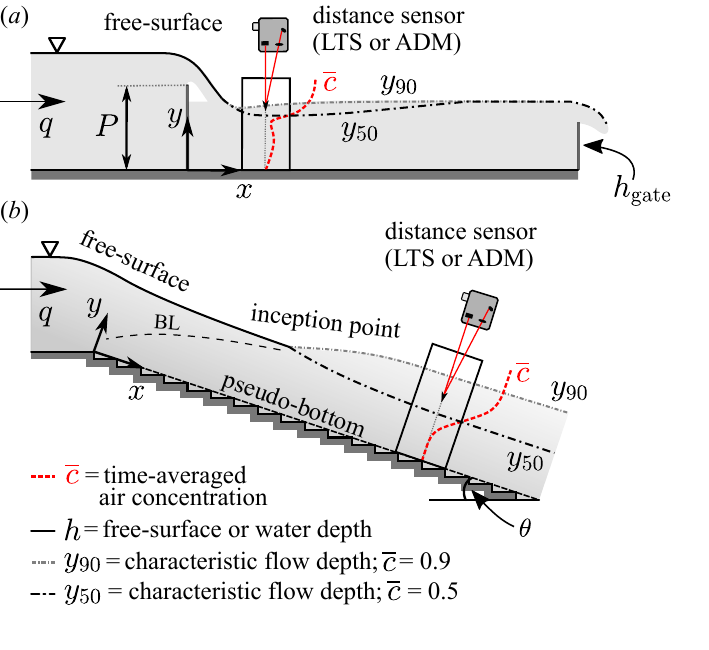}
\vspace{-0.3cm}
\caption{Experimental setup:
(\textit{a}) submerged hydraulic jump downstream of a sharp crested weir at UNSW Canberra; $h_\text{gate} = $ height of downstream sluice gate;  (\textit{b}) aerated skimming flow down a stepped spillway at FH Aachen with $\theta = $ chute angle and BL = boundary layer.} 
\label{fig:experimentalsetup}
\end{figure*}

We conducted experimental testing at two different laboratories, considering two different types of highly aerated turbulent air-water flows. A submerged hydraulic \textcolor{black}{jump} downstream of a  sharp crested weir (Fig. \ref{fig:experimentalsetup}\textit{a}) was investigated at the Hydraulics Laboratory at UNSW Canberra. The weir had a height of $P=0.4$~m and was installed within a 0.5~m wide and 9~m long hydraulic flume. The specific discharge of $q = 0.11$~m$^2$/s was controlled and measured by a variable drive centrifugal pump and an inductive flow meter, respectively. Remote and intrusive measurements were conducted along the hydraulic jump at $x=0.63P$, $0.75P$ and $1.0P$, see Table \ref{table:conditions}.

\vspace{-0.2cm}

Stepped spillway flows (Fig. \ref{fig:experimentalsetup}\textit{b}) were investigated at the Hydraulics Laboratory of FH Aachen. The stepped spillway flume had a slope angle of 26.6$^\circ$, a total drop height of 1.74 m, a constant step height of $s =$ 0.06 m, and a width of 0.5 m. Remote and intrusive measurements were carried out at step edge 21 for three different specific discharges $q=0.07$, $0.11$, and $0.15$~m$^2$/s (Table \ref{table:conditions}), which were controlled by a frequency-regulated pump and measured with an in-pipe inductive flow meter. The experimental conditions involved skimming flow for all test configurations \cite{Chanson2002}. 

\renewcommand*{\arraystretch}{1}
\begin{table}[h!]
\begin{center}
\caption{Experimental flow conditions and measurement locations; deployed measurement instrumentation included laser triangulation sensors (LTS), acoustic displacement meters (ADM), and dual-tip phase-detection conductivity probes (PD)}
\begin{tabular}{c c c c c c c}
\toprule
Parameter & submerged HJ & stepped spillway  \\
\midrule
$q$ (m$^2$/s) & 0.11 & 0.07, 0.11, 0.15 \\
$P$ (m) & 0.4 & - \\
$h_\text{gate}/P$ (-) & 0.5 & -\\
$x/P$ (-) &  0.63, 0.75, 1.0 & - \\
$\theta$  ($^\circ$) & 0 & 26.6\\
$s$ (m) & - & 0.06\\
step edge (-) & - & 21\\
Sensors (-) & LTS, ADM, PD & LTS, ADM, PD\\
\bottomrule
\end{tabular}
\label{table:conditions}
\end{center}
\end{table}

In addition to the LTS (see $\S$ \ref{sec:laser}), we deployed acoustic displacement meters (ADM) and dual-tip phase-detection conductivity probes (PD) for free-surface detection and intrusive measurements of air-water flow properties, respectively. All instruments were mounted at identical streamwise locations as per Table \ref{table:conditions}. We used Microsonic mic25+/IU/TC sensors with measuring range of 65 to 350 mm and a response time of 64 ms, which were sampled at 300 Hz for a sampling time of 600 s. The ADMs were installed around 300 mm above the (pseudo-) bottoms of the flumes, and referenced to the particular zero-level. To avoid wetting of the membrane, the ADMs were protected by surrounding cones. We used the Robust Outlier Cutoff \cite[ROC]{Valero2020} for outlier removal, and we obtained air concentrations from ADM measurements using the approach of \citeN{Cui2022}. 

The deployed phase-detection probe at FH Aachen followed current state-of-the-art design with identical leading and trailing tips with inner electrodes of platinum wire (diameter $0.13$ mm) and outer needle tips made of stainless steel (diameter $0.4$ mm). The UNSW Canberra probe was a new development. Here, inner and outer electrodes of both tips were printed on a fine circuit board with thickness of 0.2 mm. For both probes, the two probe tips were positioned side-by-side, separated by $\Delta x = 4.85$ mm (FH Aachen) and $\Delta x = 5$ mm (UNSW Canberra), as well as $\Delta z = 0.7$ mm (FH Aachen) and $\Delta z =1$  mm (UNSW Canberra)  in longitudinal and transverse directions. The measurement of vertical profiles was automated with linear motion control, and air-water signals were recorded for 45 s at sampling rates of 20 kHz (UNSW Canberra) and 100 kHz (FH Aachen) using high-speed data acquisition systems. We post-processed the raw data with a 50 \% single threshold technique \cite{Cartellier1991}, which provided time-averaged air concentrations ($\overline{c}$) and interface frequencies ($F$) at each measurement location, as well as interfacial velocities ($\overline{u}$) and their fluctuations using the adaptive window cross-correlation (AWCC) technique \cite{Kramer19AWCC,Kramer2020}.

\begin{figure}[h!]
\centering
\includegraphics{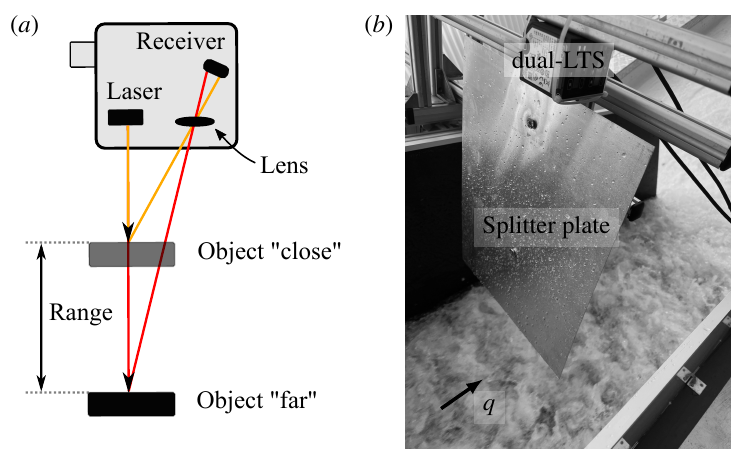}
\caption{Optical distance measurement principles: (\textit{a}) laser triangulation principle; (\textit{b}) dual-LTS arrangement with splitter plate above stepped spillway; \textcolor{black}{note that the two LTS were mounted next to each other, without requiring a specific casing.}}
\label{Fig2}
\end{figure} 

\newpage
\section{Laser distance sensing}
\label{sec:laser}

In the current study, we deployed LTS of type Sick OD2000-7002T15 for all experiments. The sensor uses a triangulation principle by projecting a light spot on the detectable object (here: aerated surface), whose reflection is detected by a light-sensitive receiver (Fig.~\ref{Fig2}\textit{a}). Depending on the distance between the sensor and object, the laser beam is reflected with a different angle, leading to a different location on the receiver element (Fig.~\ref{Fig2}\textit{a}). 

The LTS measuring range is between 200 and 1,200 mm, with a response time of 0.533 ms. The class 2 laser emits a visible red light beam with 655 nm wavelength, characterised by a typical light spot diameter of 1 mm at 700 mm distance. We operated the Sick OD2000-7002T15 sensors with a voltage output, while sample rate and time were fixed at 19.2 kHz and 600 s, respectively. Further, we disabled all integrated filter options; filtering was performed after recording, as detailed in $\S$ \ref{sec:filtering}. For both experimfental setups, the LTS were installed approximately 700 mm above the \mbox{(pseudo-)} bottoms of the flumes, at streamwise locations as per Table \ref{table:conditions}. To extract flow velocities and turbulence properties, a setup with two identical LTS was arranged for the stepped spillway tests (Fig.~\ref{Fig2}\textit{b}). The receivers of the LTS were separated by approximately 27 mm in streamwise distance, which was due to the case dimensions. Further, we installed a splitter plate to avoid that receivers captured the light reflection of the adjacent LTS (Fig.~\ref{Fig2}\textit{b}).

\subsection{Signal characterisation and filtering}
\label{sec:filtering}
\begin{figure}[h!]
\centering
\includegraphics{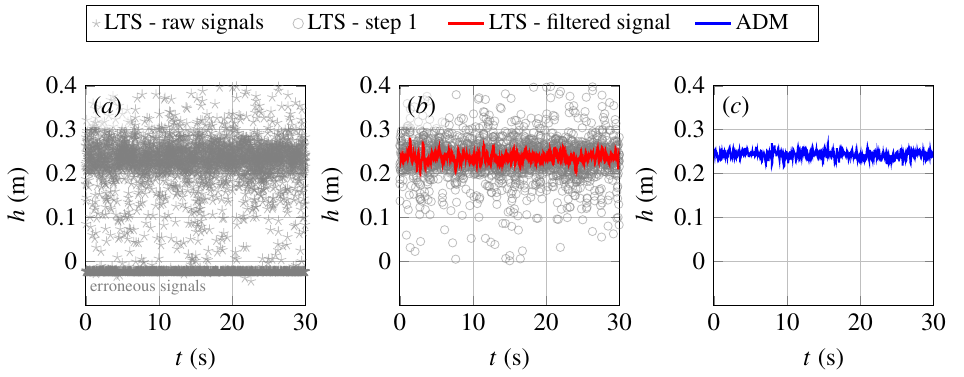}
\caption{LTS signal processing, exemplified for the submerged HJ at $x/P = 0.63$: (\textit{a}) LTS raw signal; note that the data have been down-sampled to 150 Hz for visualization; (\textit{b}) LTS signals after filtering step 1 (gray circles), and after steps 2 and 3 (red line);  (\textit{c}) ADM signal for comparison \textcolor{black}{(not simultaneously recorded)}.} 
\label{Fig3}
\end{figure}

Figure \ref{Fig3}\textit{a} shows exemplary raw signals [$h(t)$] of the LTS for the submergend HJ at $x/P = 0.63$. Erroneous measurements imply that no reflection was detected by the receiver; these data are displayed as user defined distance outside the LTS measuring range. 

For filtering purposes, we implement the following steps:
\begin{enumerate}
 \setlength\itemsep{0em}
    \item First, we replace erroneous data, including their surrounding $\pm 10$ measurements, with NaN values, see Figs. \ref{Fig3}\textit{a,b} for an illustration,
    \item In a second step, we apply a moving median filter. The number of weighted data points was found through detailed sensitivity analyses (Appendix \ref{Appendix}),
    \item For outlier removal, we apply a simplified version of the despiking method proposed by \citeN{Goring02}, as modified by \citeN{Wahl03}. This method, called Robust Outlier Cutoff \cite[ROC]{Valero2020}, is performed iteratively until no more outliers are rejected. Outliers are accounted as NaN data.
\end{enumerate}

\subsection{Instantaneous air concentration}
\begin{figure}[h!]
\centering
\includegraphics{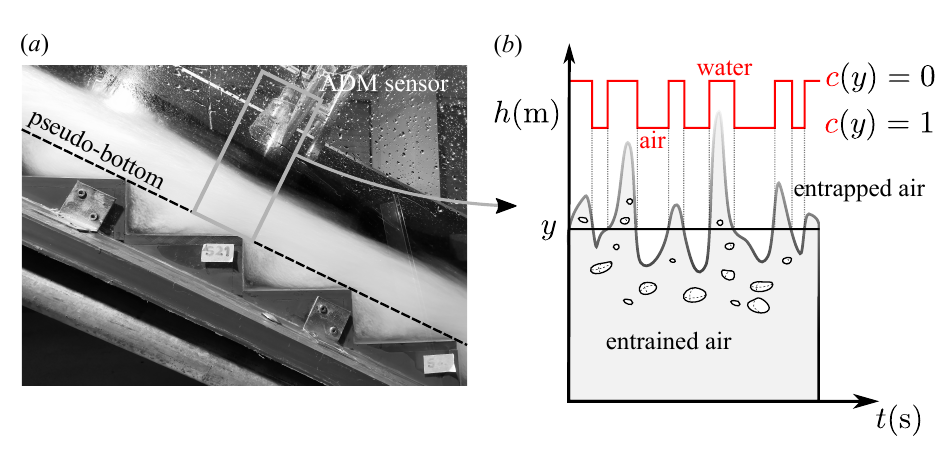}
\caption{Conversion of air-water surface signals into a time series of instantaneous entrapped air concentrations: (\textit{a}) time-averaged capture of the flow down the stepped spillway with $q = 0.11$ m$^2$/s; (\textit{b}) visual representation of Eq. (\ref{eq:convert}).}
\label{Fig4}
\end{figure}

The LTS recorded time-varying air-water surface profiles $h(t)$, which we converted into instantaneous entrapped air concentrations ($c_\text{trap}$) for different bed-normal elevations ($y$) using
\begin{equation}
c_\text{trap}(y,t) = \begin{cases} 
0, & \text{if }  h(t) \ge y, \\ 
1, & \text{if } h(t) < y,
\end{cases}
\label{eq:convert}
\end{equation}
where we assumed that (i) the laser beam was reflected at the wavy air-water interface \cite{Rak2023}, and that (ii) entrained air bubbles did not contribute to the signal, see Fig. \ref{Fig4} for illustration.

As such, Eq. (\ref{eq:convert}) allows to convert air-water elevation data into time-series of instantaneous air concentrations. Subsequent processing is similar to the analysis of phase-detection probe signals, see \citeN{FELDER201566} and \citeN{Kramer2020}, thereby providing chord times, interface count rates ($F$), and time-averaged air concentrations, the latter being calculated as
\begin{equation}
    \overline{c}_\text{trap}(y) = \frac{1}{T} \int_{t=0}^{t=T} c(y,t)_\text{trap} \,\text{d}t, \  
    \label{eq:ctimeaverage}
\end{equation}
where $T$ is the sampling duration, which was significantly larger than the time scale of turbulent processes.

\subsection{Interfacial velocities and turbulence}
Velocities of the flows down the stepped spillway were estimated using instantaneous air concentrations of the dual-LTS arrangement [Eq. (\ref{eq:convert})], further requiring recordings from both, leading and  trailing sensor, each at a particular elevation $y$. Here, we applied the adaptive window cross-correlation (AWCC) technique following  \citeN{Kramer19AWCC}, \citeN{Kramer2020}, and \citeN{Kramer20Practicescomment}. The AWCC technique segments the time series of instantaneous air concentrations (c.f., Fig.~\ref{eq:convert}) into small windows, each containing a reduced number of phase changes. Subsequently, pseudo-instantaneous interfacial velocities and turbulence levels were estimated via cross-correlation analysis, adhering to several cross-correlation quality criteria, such as the cross-correlation coefficient and the secondary peak ratio \cite{Kramer2020}.

\section{Results}
\subsection{Air concentration and interface frequencies}
\label{sec:airconcentration}

\begin{figure*}[h!]
\centering
\includegraphics{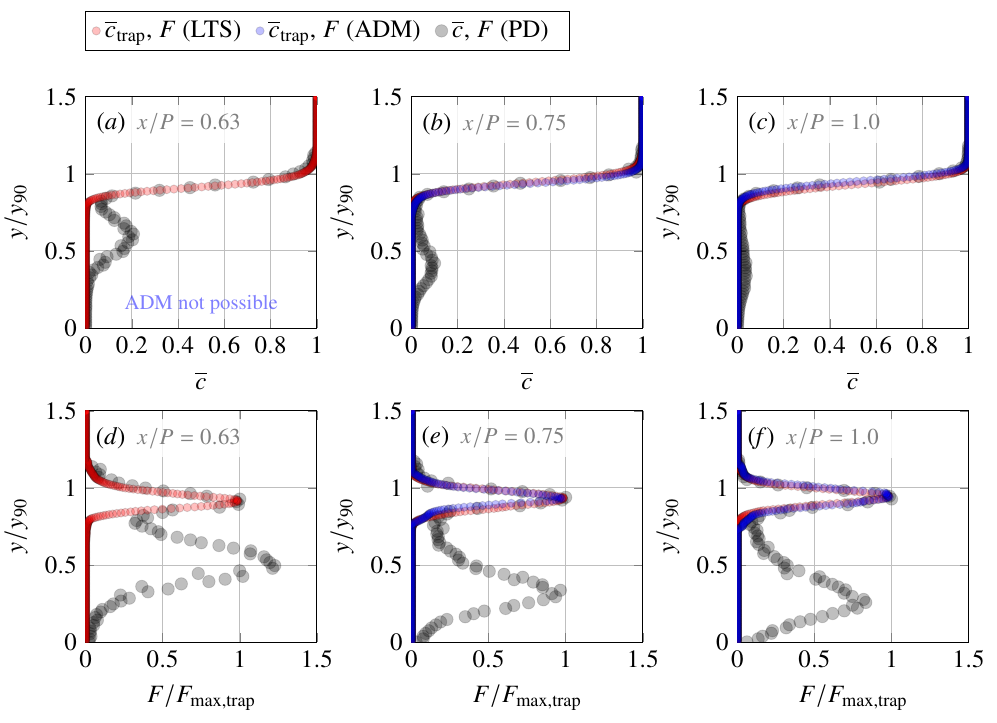}
\caption{Comparison between phase-detection probe (PD), laser triangulation (LTS), and acoustic displacement meter (ADM) measurements for the submerged HJ: (\textit{a} to \textit{c}) air concentrations;
 (\textit{d} to \textit{f}) interface frequencies; \textcolor{black}{note that all data is normalized with $y_{90}$ obtained by the intrusive PD.}}
\label{Fig5}
\end{figure*}

Here, we present our measurements of the submerged HJ by comparing the results of the three deployed flow measurement instruments, including phase-detection probe (PD), laser triangulation sensor (LTS), and acoustic displacement meter (ADM), see Fig. \ref{Fig5}. We note that ADM measurements closest to the jump toe, at $x/P = 0.63$, were not possible, due to the membrane of the sensor being exposed to splashing, which resulted in non-physical erroneous readings.

The air concentrations measured with the PD (Figures \ref{Fig5}\textit{a} to \textit{c}) were similar to classical air concentrations of hydraulic jumps, which have been well documented in literature, see \citeN{Wang2014}, amongst others. When comparing PD measurements with measurements of the LTS and the ADM, it becomes clear that the latter two sensors did measure entrapped air within the wavy free-surface layer, while they were not able to detect entrained air bubbles. This is also supported by the measurement results of the interface count rates (Figures \ref{Fig5}\textit{d} to \textit{f}), showing an excellent agreement of the three measurement instruments for the wavy free-surface region. We note that the interface count rates were normalised with their respective maximum $F_\text{max,trap}$ (Figures \ref{Fig5}\textit{d} to \textit{f}), while absolute $F$-values were highest for the PD, similar to \citeN{Cui2022}. Overall, we conclude that both flow measurement instruments, the LTS and the ADM, were able to measure entrapped air, as exemplified in Fig. \ref{Fig4}\textit{b}. Based on this finding, we had already adopted $c_\text{trap}$ in the definition of Eqns. (\ref{eq:convert}) and (\ref{eq:ctimeaverage}).

\begin{figure*}[h!]
\centering
\includegraphics{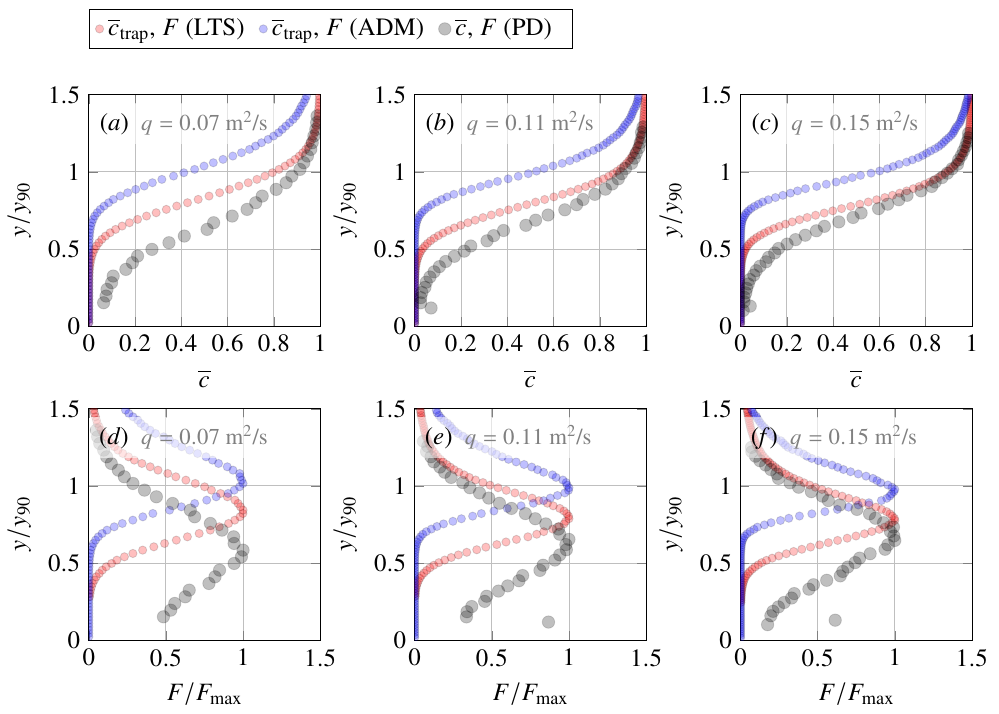}
\caption{Comparison between phase-detection probe (PD), laser triangulation (LTS), and acoustic displacement meter (ADM) measurements for the stepped spillway: (\textit{a} to \textit{c}) air concentrations;
 (\textit{d} to \textit{f}) interface frequencies; \textcolor{black}{note that all data is normalized with $y_{90}$ obtained by the intrusive PD.}}
 \label{Fig6}
\end{figure*}

For the stepped spillway experiment, we find notable differences in the detected air concentrations of the three measurement instruments (Figures \ref{Fig6}\textit{a} to \textit{c}). First, the differences between the remote sensing instrumentation (LTS and ADM) and the intrusive PD are expected, as the PD is detecting the total conveyed air, while LTS and ADM are anticipated to measure entrapped air. Second, the differences between the two remote sensing instruments (LTS and ADM) are likely a result of the sensor's different beam sizes, \textcolor{black}{in combination with the respective structure of the flow in the aerated free-surface layer, which was dominated by waves (submerged hydraulic jump), and by a combination of waves, entrained air bubbles, and ejected droplets (stepped spillway), see also \citeN{Kramer2024}.} In our experiment, the beam diameter of the ADM (40 mm at 300 mm distance; 300 mm corresponds to flume bottom) is significantly larger than the beam diameter of the LTS (1 mm at 700 mm distance; 700 mm corresponds to flume bottom), which causes a bias towards foreground flow features; as such, the LTS sensor is anticipated to be more accurate. For larger sensors, more particles are detected, leading to an increase in $y_{90}$. We note that a much more pronounced self-similarity is found in the turbulent wavy layer for all sensors if the particular $y_{90}$ for each instrument is considered (not illustrated here). Additionally, the sampling bias of the ADM beams becomes relevant for the spillway tests, as the free-surface roughness is governed by highly three-dimensional processes \cite{Zhang2018b}.

Similar conclusions may be drawn from the resulting interface frequencies in \ref{Fig6}\textit{d} to \textit{f}. The data follows a similar trend for all sensors, while again an upward shift of the interface frequency distributions is noted. Highest deviation is found again for the ADM while LTS data is much closer to the intrusive probe. For higher discharges, deviations become smaller and the normalized frequencies obtained in the wavy layer with the LTS \textcolor{black}{are} very close to the PD data. Discrepancies between ADM and PD are smaller for $q = 0.15$~m$^2$/s as well.

\subsection{Interfacial velocities and turbulence}
\label{sec:velocities}

In this section, we show that the dual-LTS arrangement may be considered as a suitable remote sensing approach for estimation of both, velocity and Reynolds stress distributions in the turbulent wavy layer of highly aerated flows. To obtain pseudo-instantaneous interfacial velocity time series, we process the instantaneous entrapped air concentrations of the LTS (Eq.~\ref{eq:convert}) as well as the PD signals using the AWCC technique \cite{Kramer19AWCC}, following best practice guidelines proposed by \citeN{Kramer2020}. The selection of the number of particles/interface pairs is detailed in Appendix \ref{Appendix2}.

\begin{figure*}[h!]
\centering
\includegraphics{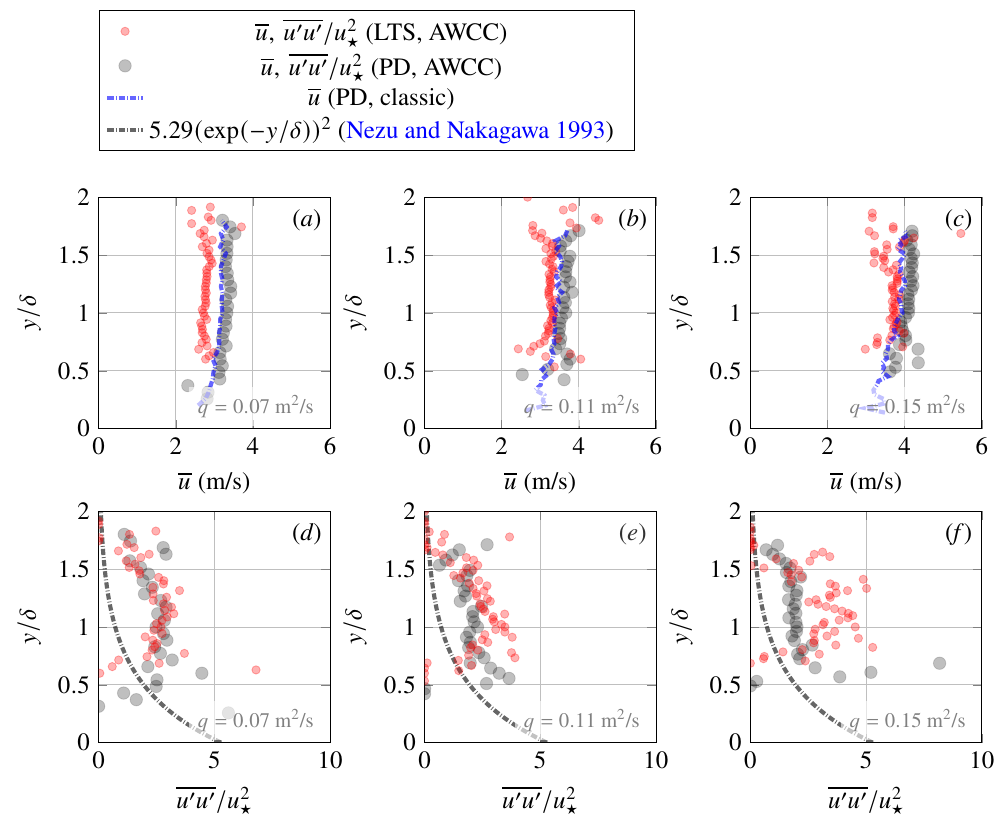}
\caption{Mean velocities and normal turbulent stresses - comparison between PD and LTS  for the stepped spillway at SE21: (\textit{a}, \textit{d}) $q = 0.07$ m$^2/s$; (\textit{b}, \textit{e}) $q = 0.11$ m$^2/s$; (\textit{c}, \textit{f}) $q = 0.15$ m$^2/s$.}
\label{fig7}
\end{figure*}

Resulting velocity profiles from both flow measurement instruments (PD and LTS) are presented in Figs. \ref{fig7}\textit{a} to \textit{c}, including a comparison with mean PD velocities obtained by cross-correlation of the full signals, denoted as classic approach. In Fig. \ref{fig7}, the bed-normal coordinate is normalised with the boundary layer thickness $\delta$,  defined as the bed-normal distance from the pseudo-bottom where the velocity reached 99\% of its maximum value. A good agreement of PD data (AWCC and classic) is noted for absolute mean velocities as well as the shape of their distributions, \textcolor{black}{while the AWCC data showed some scatter close to the pseudo-bottom, which is due to the lower data yield, previously discussed in} \citeN{Kramer20Practicescomment}. Despite showing an overall reasonable agreement, mean velocities of the LTS are slightly lower than PD velocities (Figs. \ref{fig7}\textit{a} to \textit{c}), which may result from the larger separation distance of the LTS (27 mm) when compared to the PD (5 mm). We anticipate that this larger distance can impact the suitability of some signal segments for velocity estimation, as interface pairs may not be detected by both sensors due to the three-dimensional nature of the flow. Further, we note that the laser beam cannot penetrate into the turbulent boundary layer, and as such, LTS data is missing closer to the pseudo-bottom of the stepped spillway (Figs. \ref{fig7}\textit{a} to \textit{c}). 

To demonstrate the LTS capability to capture velocity fluctuations, we present dimensionless Reynolds stress distributions $\overline{u'u'}/{u_{\star}^2}$ in Figs. \ref{fig7}\textit{d} to \textit{f}. Here, we obtained shear velocities using $u_{\star} =  \sqrt{d_\text{eq} \, g \sin{\theta}}$, with $g$ being the gravitational acceleration and $d_\text{eq}$ the equivalent clear water flow depth 
\begin{equation}
d_\text{eq} = \int_{y=0}^{y_{90}} (1-\overline{c}) \, \text{d}y.  
\end{equation}

Overall, we observe a \textcolor{black}{fairly} good agreement in dimensionless Reynolds stresses between LTS and PD (Figs. \ref{fig7}\textit{d} to \textit{f}), while we acknowledge that the selection of AWCC processing parameters, as per Appendix \ref{Appendix2}, has some effects on obtained turbulent quantities, see discussion in \citeN{Kramer2020}. Lastly, we show that obtained normal stresses are in good agreement with 
semi-empirical open-channel flow relationships from \citeN{Nezu1993} (Figs. \ref{fig7}\textit{d} to \textit{f}). The self-similarity of turbulence statistics as
well as the applicability of semi-empirical open-channel flow relations has previously been reported for aerated \cite{Wang2022chute,Kramer2023stresses} and for  nonaerated \cite{Bayon2018,Toro2016,Toro2017} free-surface flows over triangular cavities. 

\section{Discussion: Two-state superposition}
\label{sec:discussion}
To compare present air concentration results with literature data, we characterize our stepped spillway measurements using the two-state superposition principle described in \citeN{Kramer2023} and \citeN{Kramer2024}. The two-state superposition approach assumes a fluctuating interface that separates the turbulent boundary layer (TBL) and the turbulent wavy layer (TWL). A convolution of the two states with a Gaussian interface probability led to the following expression of the total conveyed air concentration \cite{Kramer2024}
\begin{equation}
\overline{c} = \left(\overline{c}_\text{trap} +   \overline{c}_\text{ent}\right)  \Gamma +  \overline{c}_{\text{TBL}}(1-\Gamma), 
\label{eq:voidfractionfinal1}
\end{equation}
with
\begin{equation}
\Gamma = \frac{1}{2} \left( 1+\erf \left(\frac{y - y_\star }{ \sqrt{2} \sigma_\star}  \right)  \right).
\label{eq:gaussianerr}
\end{equation}
where  $\erf$ is the Gaussian error function, $y$ is the bed-normal coordinate, $y_\star$ is the time-averaged interface position, and $\sigma_\star$ is its standard deviation. The entrapped and the entrained air concentrations of the TWL are defined as \cite{Kramer2024}

\begin{equation}
\overline{c}_\text{trap}
= \frac{1}{2} \left(1 + \erf  \left( \frac{y - y_{50_\text{trap}}}{\sqrt{2} \mathcal{H}_\text{trap}} \right)  \right),
\label{entrappedair}
\end{equation}

\begin{eqnarray}
\overline{c}_\text{ent} = \frac{1}{2}  \left( \erf  \left( \frac{y - y_{50}}{\sqrt{2} \mathcal{H}} \right)  -  \erf  \left( \frac{y - y_{50_\text{trap}}}{ \sqrt{2} \, \mathcal{H}_\text{trap} } \right) \right),  \label{entrainedair}
\end{eqnarray}
with $y_{50_\text{trap}}$ being the free-surface level, $\mathcal{H}_\text{trap}$ the root-mean-square wave height, $y_{50}$ the characteristic mixture flow depth $y_{50} = y(\overline{c}=0.5)$, and $\mathcal{H}$ is a characteristic length scale proportional to the thickness of the wavy surface-layer. The air concentration of the TBL is written as \cite{Kramer2023}

\begin{equation}
\overline{c}_\text{TBL} = 
\begin{cases}
\overline{c}_{\delta/2}  \left(\frac{y}{\delta-y} \right)^{\beta}, & y \leq \delta/2,
\label{eq:voidfraction1}\\
\vphantom{\left(\frac{\frac{\delta}{y_\star}-1}{\frac{\delta}{y}-1} \right)^{\frac{\overline{v}_r S_c}{\kappa u_*}}}
\overline{c}_{\delta/2} \, \exp \left(\frac{4\beta}{\delta} \left(y - \frac{\delta}{2} \right)   \right), \quad & y > \delta/2, \\
\end{cases}
\end{equation}
where $\overline{c}_{\delta/2}$ is the air concentration at half the boundary layer thickness ($\delta$), and $\beta$ is the Rouse number for air bubbles in water. Note that the two-state formulation [Eq. (\ref{eq:voidfractionfinal1})] has been successfully validated against more than 500 air concentration data-sets from literature, hinting at universal applicability. 

\begin{figure*}[h!]
\centering
\includegraphics{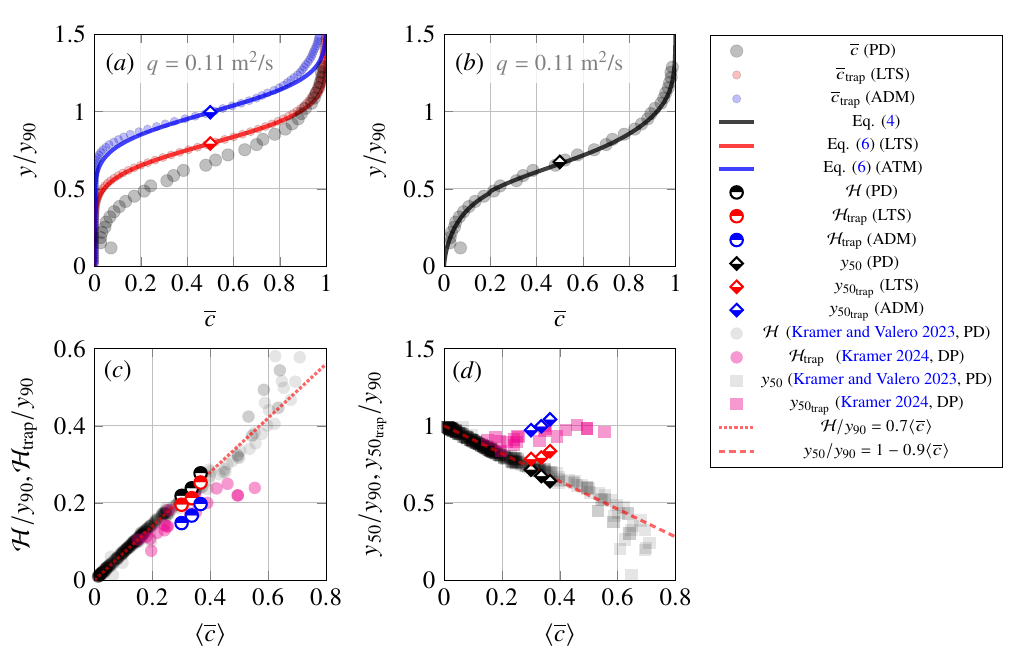}
\label{fig:discussion} 
\caption{Application of the two-state superposition approach and comparison with literature data: (\textit{a}) characterization of entrapped air; (\textit{b}) characterization of total conveyed air; (\textit{c}) root mean square wave height $\mathcal{H}_\text{trap}$ and characteristic length scale $\mathcal{H}$ versus mean air concentration;  (\textit{d}) free-surface level $y_{50_\text{trap}}$ and characteristic depth $y_{50}$ versus mean air concentration; DP = dipping probe.}
\label{Fig8}
\end{figure*}

Figures \ref{Fig8}\textit{a},\textit{b} demonstrate the application of the two-state superposition approach to the spillway data with $q = 0.11$ m$^2$/s. For conciseness, we only plot Eqns. (\ref{entrappedair}) and (\ref{eq:voidfractionfinal1}) with corresponding $y_{50_\text{trap}}$ and $y_{50}$. It is seen that Eq. (\ref{entrappedair}) is well in accordance with our measurements of entrapped air (LTS and ADM), while the complete two-state superposition formulation excellently reflects our total conveyed air (PD) measurements (Fig. \ref{Fig8}\textit{b}). When plotted against the depth-averaged (mean) air concentration $\langle \overline{c} \rangle$, the physical parameters \textcolor{black}{$\mathcal{H}$} (black half circles) and \textcolor{black}{$y_{50}$} (black half squares), both evaluated from our PD measurements, are also well in accordance with previous phase-detection literature data from \citeN{Kramer2023} (Figures \ref{Fig8}\textit{c},\textit{d}). 

Related to the measurement of entrapped air, we find that detected root-mean-square wave heights $\mathcal{H}_\text{trap}$ of the LTS (red half circles) and the ADM (blue half circles) somewhat scatter around previous dipping probe data (DP, \textcolor{black}{pink} circles) from \citeN{Killen1968}, as reanalyzed by \citeN{Kramer2024}, see Fig. \ref{Fig8}\textit{c}. In contrast, free-surface levels $y_{50_\text{trap}}$ were lower/higher for the LTS (red half squares) and the ADM (blue half squares) when compared to the dipping probe data (DP, 
\textcolor{black}{pink} squares) of \citeN{Killen1968} (Fig. \ref{Fig8}\textit{d}). There are several explanations for these discrepancies, which include foreground bias of the ADM \cite{Zhang2018b}, as well as quasi-diffusive reflection of the LTS's laser beam at entrained bubbles or ejected droplets \cite{Rak2023}. Further, the dipping probe of \citeN{Killen1968} is a unique development with potential bias due to its non-finite size \textcolor{black}{of 0.6 cm}. It is noteworthy mentioning that no other researchers have deployed such a probe since. 

At this stage, we would like to point out that LTS and ADM were in excellent agreement for the submerged HJ (Fig. \ref{Fig5}), while a conclusive statement on the discrepancies for the stepped spillway related to $y_{50_\text{trap}}$ (Fig. \ref{Fig8}\textit{d}) is difficult due to several  potential causes. Nonetheless, we like to encourage future research in this challenging field, as a better measurement and description of physical mechanism/processes in highly aerated flows will be beneficial for future implementation into numerical simulation environments. 

\section{Conclusion}
Remote sensing of air-water flow properties in self-aerated flows is anticipated to step-change our design approach to hydraulic infrastructure. Despite recent progress in this interesting and challenging field, seminal research questions, such as ``what do we measure'' and ``what can we measure'' with remote sensing technology, remain unanswered. In this study, we investigated the application of laser triangulation sensors (LTS) to a submerged hydraulic jump and a stepped spillway. For a comparative assessment, we also deployed acoustic displacement meters (ADM) and intrusive dual-tip phase-detection probes (PD).  

To compare the distance sensing results of the LTS and the ADM with intrusive PD measurements, we present a novel conversion algorithm, allowing us to translate time-varying air-water surface measurements into instantaneous entrapped air concentrations. Subsequently, we analysed these time series of instantaneous air concentrations using well-established signal processing methods, developed for phase-detection probe outputs. As such, we are able to remotely extract basic air-water flow properties, including elevations, air concentrations, chord times, and interface frequencies. In addition, a dual-LTS setup enabled us, for the first time, to remotely measure interfacial velocity profiles and associated turbulence levels of the flows down the stepped spillway.  

\textcolor{black}{On the question ``what do we measure''}, our results show that LTS and ADM were able to reliably measure entrapped air concentrations and interface frequencies of the submerged hydraulic jump, while entrained air bubbles were not detected by the remote sensing technologies. For the flow down the stepped spillway, we find differences in measured air concentrations between the LTS and the ADM. This measurement bias is likely a result of the sensor's beam sizes, and we anticipate that the smaller beam of the LTS is more accurate, while we acknowledge that LTS measurements could be subject to quasi-diffusive reflection at entrained air bubbles/ejected droplets. While more systematic tests are required to shed light on these different biases, the overall suitability of remote sensing technology for measuring air-water surface levels and entrapped air concentrations is deemed accurate and promising. Our remotely measured interfacial velocities and turbulence levels were in good agreement with intrusive phase-detection probe measurements, providing a first proof-of-concept of the dual-LTS setup.

Overall, our research shows that the interpretation of remotely sensed distance signals of highly aerated flows can be significantly expanded, beyond the traditional presentation of time-averaged elevations and their standard deviations, \textcolor{black}{thereby providing answers to the question ``what can we measure''}. With relative ease, remote distance signals can be converted into time series of instantaneous air concentrations, enabling the application of a variety of existing data analysis techniques. There is a wealth of information to be explored, which we hope will be used to gain more insights into the underlying physical processes of complex, three-dimensional real-world air-water flows. 

\appendix
\section{Filter length}
\label{Appendix}
For the second filtering step, we smooth the remotely sensed LTS signals using a moving median filter, whose filter length needs to be determined. We hypothesize that the air concentration at $y_{99}$, i.e., the mixture flow depth where $\overline{c} = 0.99$, consists of entrapped air only. Thus, the air concentration measured intrusively (PD) and non-intrusively (LTS) must be identical for $y \geq y_{99}$, which further implies that the entrained air concentration of the wavy free-surface layer, calculated as \cite{Kramer2024} 
\begin{equation}
\overline{c}_\text{ent} = \overline{c}_\text{TWL} - \overline{c}_\text{trap}    
\end{equation}
approximates $\overline{c}_\text{ent} \approx 0$ for $y \geq y_{99}$, i.e., there is no entrained air in the droplet/spray region. We acknowledge that our hypothesis neglects the detection of ejected droplets with the LTS sensors, which we anticipate to have only marginal influence on our results.

\begin{figure*}[h!]
\centering
\includegraphics{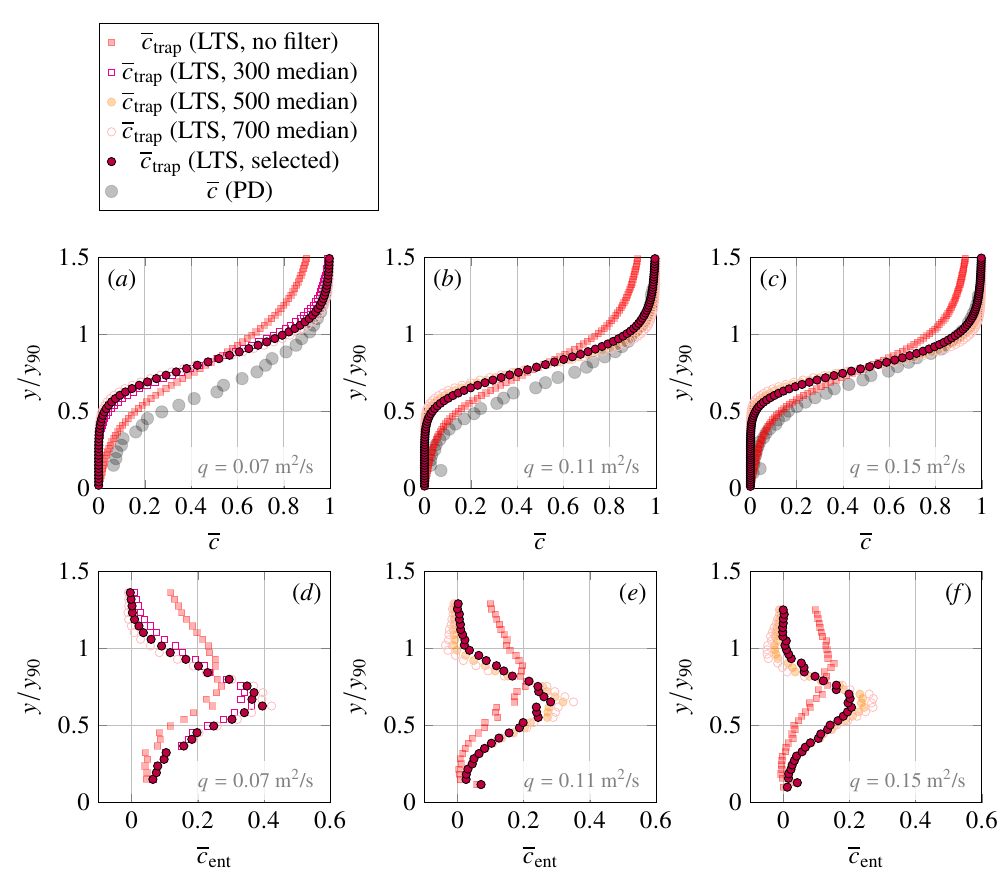}
\caption{Sensitivity analysis on the effects of median filter length on entrapped and entrained air concentration distributions, measured with leading LTS during stepped spillway experiments at SE21: (\textit{a}, \textit{d}) $q = 0.07$ m$^2/s$; (\textit{b}, \textit{e}) $q = 0.11$ m$^2/s$; (\textit{c}, \textit{f}) $q = 0.15$ m$^2/s$.}
\label{Fig9}
\end{figure*}

Figures \ref{Fig9}\textit{a} to \textit{c} show total conveyed air concentrations (PD) as well as entrapped air concentrations (LTS) for the stepped spillway experiments, where the latter were calculated using different filter lengths of 300, 500, and 700 surrounding data points. We find that a larger amount of entrapped air ($\overline{c}_\text{trap}$) is obtained for increasing filter lengths, and that the entrapped air at $y_{99}$ generally matches the conductivity probe data for a filter length of 500 data points at the selected sample rate of 19.2 kHz. However, too long filters yield negative amounts of entrained air ($\overline{c}_\text{ent}$) at approximately $y_{90}$  (Figures \ref{Fig9}\textit{d} to \textit{f}), which is physically not plausible. Therefore, we used case-sensitive filter lengths of 500 ($q=$~0.07 m$^2$/s), 300 ($q=$~0.11 m$^2$/s), and 300 ($q=$~0.15 m$^2$/s) data points, while we note that  free-surface levels $y_{50_\text{trap}}$ were independent of the selected length. 
The decreasing filter length may be explained with the more stable skimming flow regime for higher discharges. For the submerged hydraulic jump, the filter lengths were in the same order, around 500 data points, and their determination was straightforward, which is because there was a good agreement of measured total conveyed air concentrations and entrapped air concentration profiles in the wavy free-surface region (Fig. \ref{Fig5}), suggesting that the contribution of entrained air bubbles is insignificant in this region. 


\newpage
\section{AWCC signal processing parameters}
\label{Appendix2}

In this Appendix, we perform a sensitivity analysis on the selection of the AWCC processing parameters for the dual-LTS arrangement, particularly with respect to the number of particles/interface pairs $n$. We note that $n = 10$ for PD signals was selected following \citeN{Kramer2020}. Other AWCC parameters for the dual-LTS arrangement, including cross-correlation based filtering criteria, were selected in accordance with \citeN{Kramer2020} as 
\begin{equation}
\frac{R_{12,\text{max}}}{\text{SPR}^2 + 1} > 0.4, 
\end{equation}
where $R_{12,\text{max}}$ and SPR are the maximum cross-correlation coefficient and the secondary peak ratio for each window, respectively. 

\begin{figure*}[h!]
\centering
\input{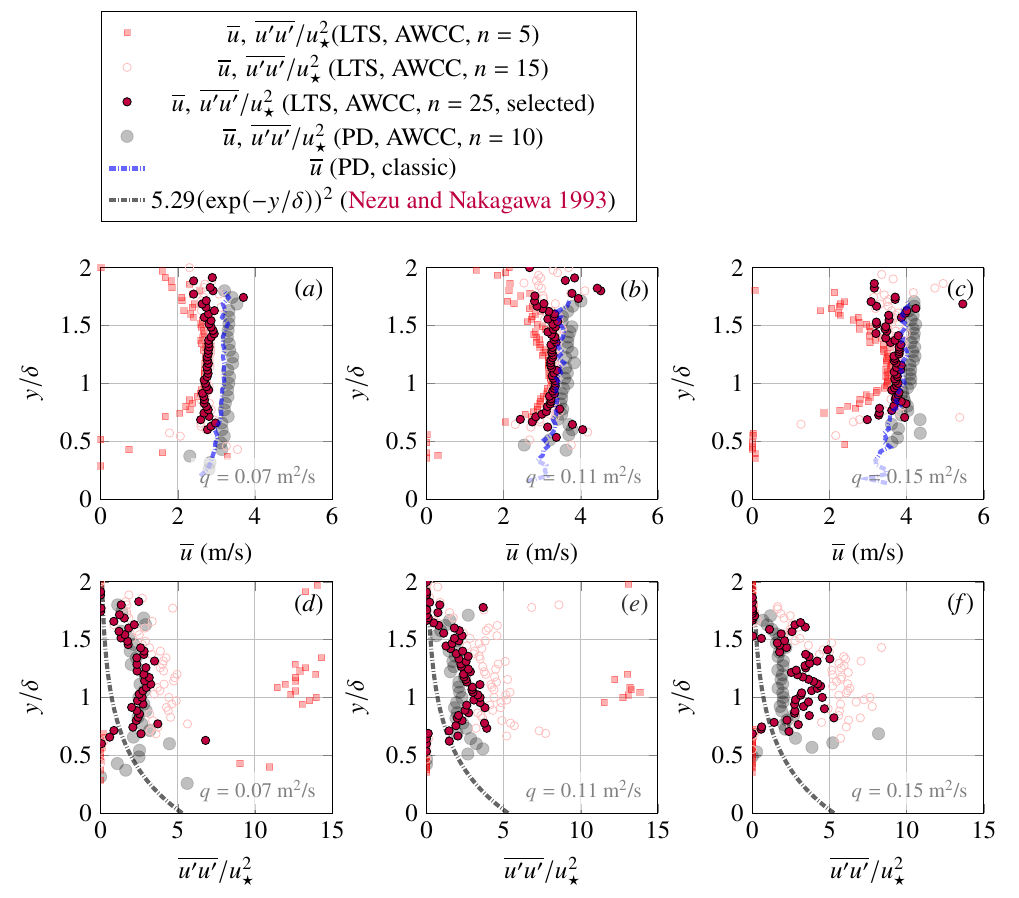}
\caption{Sensitivity analysis on the effect of number of interface pairs ($n$) on mean velocities and Reynolds stresses, measured with the dual-LTS arrangement during the stepped spillway experiments at SE21: (\textit{a}, \textit{d}) $q = 0.07$ m$^2/s$; (\textit{b}, \textit{e}) $q = 0.11$ m$^2/s$; (\textit{c}, \textit{f}) $q = 0.15$ m$^2/s$.}
\label{Fig10}
\end{figure*}

Figure \ref{Fig10} shows the results of our sensitivity analysis, where we used $n = 5, 15$, and $25$ for the LTS signals. It is seen that mean LTS velocities converge for $n \geq 15$ (Figs. \ref{Fig10}\textit{a} to \textit{c}), while LTS normal turbulent stresses are non-physical for $n =5$, further decreasing with increasing $n$ due to averaging effects (Figs. \ref{Fig10}\textit{d} to \textit{f}). Based on our comparative analysis with PD results, we find that $n = 25$ is suitable for the recorded LTS data. We acknowledge that the number of particles/interface pairs $n$ is larger for the LTS data ($n = 25$) when compared to the PD data ($n = 10$), which is likely due to the different streamwise separation distances of 27 mm (LTS) and 5 mm (PD), respectively. Despite these differences, our research provides a first proof-of-concept for the remote measurement of mean velocity profiles and Reynolds stresses in highly aerated flows. Future research using dual-LTS arrangements with separation distances similar to commonly used PD is anticipated to shed light on selection of signal processing parameters.

\section*{Data Availability Statement}
All data, models, or code that support the findings of this study are available from the corresponding author upon reasonable request.

\section*{Notation} \label{sec:notation}
\emph{The following symbols are used in this paper:}
\vspace{-0.5cm}
\begin{tabbing}
\hspace*{0cm}\=\hspace*{2cm}\=\hspace*{9.2cm}\=\kill\\
\>$c$  \> instantaneous air concentration  \> (-)\\
\>$\overline{c}$  \> air concentration (time-averaged) \> (-)\\
\>$\langle \overline{c} \rangle$  \> depth-averaged air concentration \> (-)\\
\>$F$  \> interface frequency/interface count rate \> (s$^{-1}$)\\
\>$h$  \> measured elevation (remote sensing device) \> (m)\\
\>$h_{\text{gate}}$  \> height of downstream sluice gate  \> (m)\\
\>$\mathcal{H}$  \> length-scale of the TWL \> (m)\\
\>$\mathcal{H}_\text{trap}$  \> root-mean-square wave height \> (m)\\
\>$P$  \> height of sharp crested weir \> (m)\\
\>$q$  \> specific water discharge \> (m$^2$ s$^{-1}$)\\
\>$s$  \> step height \> (m)\\
\>$t$  \> time \> (s)\\
\>$T$  \> sampling duration \> (s)\\
\>$\overline{u}$  \> streamwise velocity (time-averaged) \> (m s$^{-1}$)\\
\>$u'$  \> fluctuating part of the streamwise velocity \> (m s$^{-1}$)\\
\>$x$  \> streamwise coordinate \> (m)\\ 
\>$y$  \> coordinate normal to the channel bed \> (m)\\ 
\>$y_{50}$  \> mixture flow depth where $\overline{c} = 0.5$ \> (m)\\
\>$y_{90}$  \> mixture flow depth where $\overline{c} = 0.9$ \> (m)\\
\>$y_\star$  \> time-averaged interface position \> (m)\\ 
\>$\beta$  \> Rouse number for air bubbles in water \> (-)\\
\>$\delta$  \> boundary layer thickness \> (m)\\
\>$\Delta x$  \> probe-wise tip separation distance \> (m)\\
\>$\Delta z$  \> probe-normal tip separation distance \> (m)\\
\>$\theta$  \> chute slope \> ($^\circ$)\\
\>$\sigma_\star$  \> standard deviation of interface position \> (m)\\
\end{tabbing}
\textbf{Indices and operators}
\vspace{-0.5cm}
\begin{tabbing}
\hspace*{0cm}\=\hspace*{2cm}\=\hspace*{9.2cm}\=\kill\\
\>ADM  \> acoustic displacement meter \> \\
\>AWCC  \> adaptive window cross-correlation \> \\
\>DP  \> dipping probe \> \\
\>ent  \> entrained air \> \\
\>FS  \> free-surface \> \\
\>HJ  \> hydraulic jump \> \\
\>LIDAR  \> light detection and ranging \> \\
\>LTS  \> laser triangulation sensor \> \\
\>max  \> maximum \> \\
\>PD  \> phase-detection intrusive probe \> \\
\>ROC  \> robust outlier cutoff \> \\
\>trap  \> entrapped air \> \\
\>TBL  \> turbulent boundary layer \> \\
\>ToF  \> time of flight \> \\
\>TWL  \> turbulent wavy layer \> \\
\>$\overline{\phantom{C}}$  \> time averaging \> \\
\>$\langle \phantom{C} \rangle $ \> spatial averaging\\
\end{tabbing}

\bibliography{ascexmpl-new}

\end{document}